\documentclass{ieeeaccess}
\usepackage{cite}
\usepackage{amsmath,amssymb,amsfonts}
\usepackage{graphicx}
\usepackage{textcomp}

\usepackage{algorithm}
\usepackage{listings}
\usepackage{lipsum}
\usepackage{multirow}
\usepackage{lineno,hyperref}
\usepackage{algpseudocode}
\usepackage{booktabs}
  
\usepackage{soul}  

\usepackage{caption}
\DeclareCaptionFont{ieeeblue}{\color{accessblue}}
\DeclareCaptionLabelFormat{myformat}{\raggedright\figcapfont{\textbf{#1}\textbf{#2}}}
\def\BibTeX{{\rm B\kern-.05em{\sc i\kern-.025em b}\kern-.08em
    T\kern-.1667em\lower.7ex\hbox{E}\kern-.125emX}}
\begin{document}

\history{Received March 16, 2020, accepted March 22, 2020, date of publication March 26, 2020, date of current version April 13, 2020.}
\doi{10.1109/ACCESS.2020.2983584}

\title{Identifying Banking Transaction Descriptions via Support Vector Machine Short-Text Classification Based on a Specialized Labelled Corpus}
\author{\uppercase{Silvia Garc\'ia-M\'endez}\authorrefmark{1},
\uppercase{Milagros Fern\'andez-Gavilanes\authorrefmark{2}, Jonathan Juncal-Mart\'inez\authorrefmark{1}, Francisco J. González-Castaño\authorrefmark{1}, and \'Oscar Barba Seara\authorrefmark{3}}}
\address[1]{Information Technology Group, atlanTTic, School of Telecommunications Engineering, University of Vigo, Campus, 36310 Vigo, Spain}
\address[2]{Defense University Center, 36920 Mar\'in, Pontevedra, Spain}
\address[3]{CoinScrap Finance S.L., Cobi\'an Roffignac 2, 36002 Pontevedra, Spain}
\tfootnote{This work was partially supported by Ministerio de Econom\'ia, Industria y Competitividad under grant TEC2016-76465-C2-2-R and Xunta de Galicia under grants GRC2018/053 and ED341D-R2016/012.}

\markboth
{Silvia Garc\'ia-M\'endez \headeretal: Identifying Banking Transaction Descriptions via SVM Based on a Specialized Labelled Corpus}
{Silvia Garc\'ia-M\'endez \headeretal: Identifying Banking Transaction Descriptions via SVM Based on a Specialized Labelled Corpus}

\corresp{Corresponding author: Silvia Garc\'ia-M\'endez (e-mail: sgarcia@gti.uvigo.es).}

\begin{abstract}
Short texts are omnipresent in real-time news, social network commentaries, etc. Traditional text representation methods have been successfully applied to self-contained documents of medium size. However, information in short texts is often insufficient, due, for example, to the use of mnemonics, which makes them hard to classify. Therefore, the particularities of specific domains must be exploited. In this article we describe a novel system that combines Natural Language Processing techniques with Machine Learning algorithms to classify banking transaction descriptions for personal finance management, a problem  that was not previously considered in the literature. We trained and tested that system on a labelled dataset with real customer transactions that will be available to other researchers on request. Motivated by existing solutions in spam detection, we also propose a short text similarity detector to reduce training set size  based on the Jaccard distance. Experimental results with a two-stage classifier combining this detector with a {\sc svm} indicate a high accuracy in comparison with alternative approaches, taking into account complexity and computing time. Finally, we present a use case with a personal finance application, {\em CoinScrap}, which is available at {\em Google Play} and {\em App Store}.
\end{abstract}

\begin{keywords}
Machine Learning, Natural Language Processing, banking, personal finance management.
\end{keywords}

\titlepgskip=-15pt

\maketitle

\section{Introduction}
\label{intro}

Financial companies need to develop new strategies to keep and to expand their customer base. Their product portfolios have diversified over the years and customer behaviour has shifted from long-term loyalty to online interaction. 

The fierce competition between banks has led to a growing need to convert customer data -- which include short-text banking transaction ({\sc bt}) descriptions -- into information relevant for decision making.

Data mining has been successfully applied to finance in various ways: identifying likely candidates for loan disbursement \cite{derby2003data} and product acceptance \cite{ngai2009application}; characterizing  product segments \cite{Hu05}; and analysing customer attrition and retention \cite{IslamH15}. However, to the best of our knowledge the problem of automatic classification of short-text {\sc bt} descriptions (according to a predefined set of labels) has not yet been tackled. 

From a broader perspective, automated text classification has become a popular research area due to the many public digital text sources available. Text classification is useful for a wide range of applications, such as web searching \cite{chekuri1997web}, opinion mining \cite{vo2015target} and event detection \cite{kumaran2004text}. Nevertheless, most text classification methods are valid for long texts. Some distinctive aspects of short texts are:

\begin{enumerate}
\item Sparsity: Short texts often have fewer than 150 words and are usually organized in few sentences. They convey very little effective information. Since sparsity affects the quality of short text semantics, traditional techniques as those used for long texts are impractical \cite{cai2016context,du2016folksonomy}, as  it is difficult to extract key features from large feature spaces for accurate classification training.
\item Real-time generation: Nowadays vast amounts of information are continuously produced in the form of short messages. Consider, for example, chat and micro-blog information and news comments, among others. They reflect reactions in real-time to outside world events and, therefore, are difficult to collect. Consequently, short-text classification methods must be highly efficient.

\item Irregularity: Short-text terminology is not standardized and vocabularies are informal or specific (in our case related to banking). 
\end{enumerate}

Two key aspects are that words are seldom repeated in a given {\sc bt} description and that few words are irrelevant. The level of significance of a word cannot be simply determined by its repetition within the text. However, for the same reasons, short texts are less noisy than long texts.

Our proposal is based on Natural Language Processing ({\sc nlp}) and Machine Learning ({\sc ml}). It  characterizes financial short messages with features such as character and word $n$-grams,  which feed a supervised Support Vector Machine ({\sc svm}) classifier. Motivated by existing solutions in spam detection, we also propose  a short text similarity detector to reduce training set size  based on the Jaccard distance. Therefore, our proposal consists in  a two-stage classifier combining this detector with a {\sc svm}.  In any case, the sizes of short-text banking description datasets discourage the application of deep learning techniques \cite{lecun2015deep}.

The rest of this article is organized as follows. In Section~\ref{RelatedWork} we review the state of the art in short-text classification. In Section~\ref{task} we describe the classification problem.  Subsections~\ref{data-retrieval}-\ref{twostageversion} explain the modules of our system. In particular, Section \ref{twostageversion} describes the short text similarity detector to reduce training set size  based on the Jaccard distance.  Section~\ref{results} presents the experimental text corpora and evaluates our approach with real data. Section \ref{secoinscrap} cites a real world solution based on our approach. Finally, Section~\ref{conclusions} concludes the paper.

\section{Related work}
\label{RelatedWork}

\subsection{Customer analysis}
{\sc bt} data have grown considerably with the expansion of electronic banking \cite{Bhasin06}. The banking sector is well aware of the value of customer information covering demographics, leisure, wealth, insurance, financial transactions, and so on.

Several studies have been conducted on the analysis of customer attrition and retention. Some focus on aspects influencing customer choices, such as customer care, speed and quality of service, variety of services, fees, online accessibility, etc \cite{Aregbeyen11,RehmannEtal08,DihnEtAl2012}. Other studies have focused on customer churn (that is, leaving  one bank to another) \cite{Aregbeyen11, Keramati2016, SharmaP13a, ChenEtAl15}, fraud \cite{BarmanEtAl16,WestEtAl16} and even spatial distribution from transaction activity in commercial areas \cite{Yoshimura16}.

\subsection{Personal Finance Management}
Personal finance management or {\sc pfm} aggregates household bank accounts and offers users a view of their day-to-day personal finances. It involves planning and budgeting, cash flow control, investment, taxation, and insurance~\cite{vahidov2010situated}. It is becoming increasingly popular and many {\sc pfm} resources  
such as {\em BudgetBuddy}\footnote{Available at {\tt https://www.budgetbuddyaus.com.au/}.
}, {\em AccBiz}\footnote{Available at {\tt http://www.webunit.co.uk/clients/access/
index.html}.}, {\em Prosper}\footnote{Formerly available at {\tt https://www.prosper.com/}.}, {\em Finn}\footnote{Available at {\tt https://www.chase.com/personal/finnbank}.} and {\em Figo}\footnote{Available at {\tt https://www.figo.io/}.} exploit {\sc pfm} by recommending personalized  insurance products or long-term financing plans. These applications also provide budgeting and credit scoring tools to help households track their expenses and credit score.

\subsection{Open Banking European Regulation}
The European path to digitization is based on four pillars~\cite{zetzsche2019future}: (1) extensive reporting requirements to control systemic risk and change financial sector behaviour; (2) strict data protection rules; (3) open banking to enhance competition; and (4) a legislative framework for digital identification. In this line, the Second Payments Services Directive\footnote{Directive (EU) 2015/2366 of the European Parliament and of the Council of 25 November 2015 on payment services in the internal market, amending Directives 2002/65/EC, 2009/110/EC and 2013/36/EU; Regulation (EU) No 1093/2010; Repealing Directive 2007/64/EC, OJ of 23.12.2015, L 337/35.} ({\sc psd} 2) empowers customers to make their banking data available to third parties such as {\em FinTech} companies. In essence it paves the way for new banking products and services, by promoting competition without compromising security.

\subsection{Text classification}
Most existing approaches for text classification rely on simple document representations in word-oriented input spaces. Despite considerable efforts to introduce more sophisticated techniques for document representation such as those based on higher-order word statistics \cite{caropreso2001learner}, {\sc nlp} \cite{jacobs1992joining}, {\em string kernels} \cite{lodhi2002text} and word clusters \cite{baker1998distributional}, simple {\em bag-of-words} ({\sc bow}) approaches  \cite{harris1954distributional} are still popular. 

Different {\sc ml} methods, such as {\em Naive Bayes} \cite{mccallum1998comparison}, logistic regression \cite{nigam1999using} and {\sc svm}s \cite{joachims1998text} have been proposed for text classification.  In particular, linear classifiers, which are efficient, robust and easy to interpret, have been successful at sentiment analysis \cite{sebastiani2002machine}.

Diverse complex features have been added to these text classification models. Some examples  are parts-of-speech and phrase information \cite{lewis1992evaluation}, syntax integration by means of explicit features and implicit kernels \cite{post2013explicit}, and, for sentiment analysis, dependency tree features \cite{nakagawa2010dependency} and semantic composition models \cite{karo2007sentiment}. In \cite{wang2012baselines} it was shown that {\sc bow} and bigram features are more productive than much more complex features. Distributed word representations \cite{bengio2003neural, collobert2011natural, mikolov2013efficient} have  enriched discrete models for semi-supervised learning. Word embeddings have mostly been used to feed neural network models such as recursive tensor networks \cite{socher2011semi}, dynamic pooling networks \cite{kalchbrenner2014convolutional} and deep convolutional neural networks \cite{dos2014deep}. Finally,   direct learning of distributed vector representations of paragraphs and sentences for text classification 
was discussed in \cite{le2014distributed}.

As previously mentioned, unlike normal text classification, short-text classification must tackle the problem of sparsity \cite{aggarwal2012survey}. Rare and even missing words in training texts may appear in testing data. Most words only appear once in the texts that include them. Therefore, the term frequency-inverse document frequency ({\sc tf-idf}) metric is not representative. To address this issue, some researchers  enrich data contexts with information from Wikipedia \cite{banerjee2007clustering} and ontologies \cite{fodeh2011ontology}. However, this requires solid {\sc nlp}  knowledge and high-dimensional representations that may be expensive in terms of memory and computing time and, thus, inefficient for  real-time solutions. The more sophisticated approach in \cite{yin2014dirichlet} applied a Dirichlet multinomial mixture model for short-text classification. The approach in  \cite{cai2005document} clustered texts using the Locality Preserving Indexing ({\sc lpi}) algorithm. In recurrent neural network ({\sc rnn}),  textual trees are also computationally expensive \cite{lai2015recurrent}. Therefore, the design of efficient models is  still challenging. 

Two well-known methods for short-text classification are Probase Bag-of-Concepts short-text classification (Probase {\sc boc} {\sc stc}) \cite{wu2012probase} and Entity Explicit Semantic Analysis ({\sc esa}) \cite{gabrilovich2007computing}. {\sc esa} is based on semantic relation degrees \cite{han2009named,hu2009exploiting,ni2009mining} from Wikipedia. It associates all words in a Wikipedia page to the corresponding Wikipedia entry (concept) using the {\sc tf-idf} value as correlation metric and produces indexes that map each word in a short text to the concepts considered. Note that the short text may not mention the concept explicitly. {\sc esa} uses the vector representation of a short text as the input of a {\sc svm} classifier. Regarding Probase {\sc boc} {\sc stc}, it is based on the Probase knowledge base of entity relationships and other related information that Microsoft  extracted from massive Internet data using the {\em is-a} relationship. 
A key difference with {\sc esa} is that Probase is a knowledge base by itself that has been produced with an automatic extraction algorithm. However, as in the case of {\sc esa} indexing, {\em is-a} relationships may lack relevant information for short-text classification.

To the best of our knowledge  no previous research has considered short-text {\sc bt} classification. We propose a simple and efficient approach that could be easily adapted to other application domains.

\section{System description}
\label{task}

\begin{figure}[!htbp]
\centering
\includegraphics[width=0.5\textwidth]{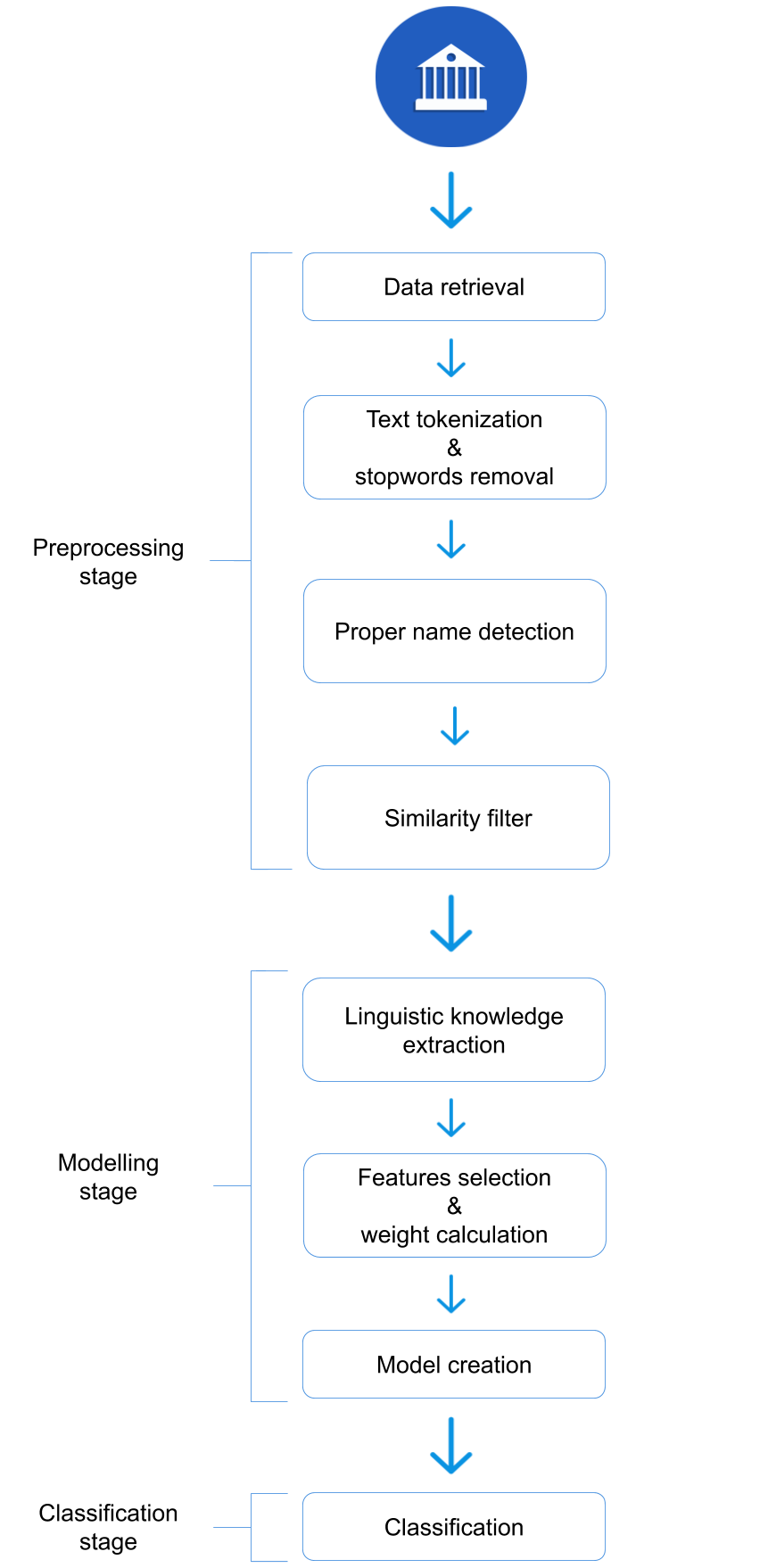}
\caption{\label{flow}System stages.}
\end{figure}

We seek to develop a simple and efficient short-text classification system by taking advantage of the particularities of {\sc bt} descriptions, with high macro-average precision, recall and $F$ measures. Our approach has three stages, as described in Figure~\ref{flow}: (1) preprocessing, (2) {\sc ml} (linguistic knowledge extraction and probabilistic model training), and (3) classification.

\subsection{Preprocessing}
\label{Preprocessing}

\subsubsection{Data retrieval}
\label{data-retrieval}

Data was retrieved with the {\em CoinScrap} coin scrapper embedded into electronic banking apps of real users, who granted us permission.  Section~\ref{sec-dataset} describes the resulting dataset.

\subsubsection{Text tokenization and stopwords}

The language of {\sc bt} descriptions is quite particular because it must be concise. The meaning of the message is condensed in few characters. In most cases verbs are totally absent. Nevertheless, {\sc bt} descriptions  may still contain useless information that may affect text classification.

First, each {\sc bt} description is split into tokens, and, in some cases, into sentences. Then meaningless words or {\em stopwords}, such as determiners and prepositions ({\em `el'}/`the', {\em `en'}/`in', {\em `entonces'}/`so', {\em `aunque'}/`although', {\em`pero'}/`but' and so on) are removed. Table~\ref{stopwordsTable} shows some stopword examples\footnote{Available at {\tt https://www.ranks.nl/stopwords/spanish}.}. Next, all punctuation marks apart from `.' and `,' are also removed.

\begin{table}[!htbp]
\centering
\caption{\label{stopwordsTable}Some examples of stopwords.}
\small
\begin{tabular}{lllll}\hline
\multicolumn{5}{c}{Stopwords}\\\hline
algún & como & incluso & poder & también \\
ambos & esta & otro & por & tras \\
ante & estar & para & primero & un\\
antes & hacer & pero & ser & uso\\
\hline
\end{tabular}
\end{table}

\subsubsection{Proper name detection}

Finally, proper names are detected using lists of names and surnames\footnote{Available at {\tt https://github.com/olea/lemarios}.} and replaced by a tag.

Taking the real {\sc bt} description {\em `Compra en supermercado Elvira Madrid 28. TARJ. :*320546'} as an example, after text tokenization, stopword removal and proper name extraction, the result is {\em`Compra \# supermercado \#PNegi\# Madrid 28. TARJ. \#320546'}. The ``\#" symbol marks the place where a word is removed. Note that each proper name is substituted by ``\#PN" followed by a  set of characters (`egi' in the example) and ``\#". Thus, a given name is always replaced by the same identifier ({\em `Elvira'} by \#PNegi\# in the example). Credit card number was always anonymized.

\subsubsection{Training sample reduction with similarity detection stage}
\label{twostageversion}

We take advantage of the fact that many {\sc bt}  descriptions are similar to reduce the size of the training set. For that purpose, we insert a similarity detector based on the Jaccard distance \cite{Seung-Seok2010} before the classifier. This is inspired by spam detection techniques that use  this  distance to seek for characteristic sentences  \cite{Harsule2016,Bajaj2017,Temma2019}. 

The similarity detector only considers the text of the descriptions.  
When the Jaccard similarity between a new labelled description and a previous  entry in our dataset exceeds 85\%, and both belong to the same category, the new description is not added to the {\sc svm} training set. Otherwise, we keep it. When the similarity between a new unlabelled description and a previous  entry  exceeds 85\%, we assign to the description the class of the entry. Otherwise, the description is passed to the {\sc svm} for classification.

\begin{figure*}[!htbp]
\centering
\includegraphics[width=0.95\textwidth]{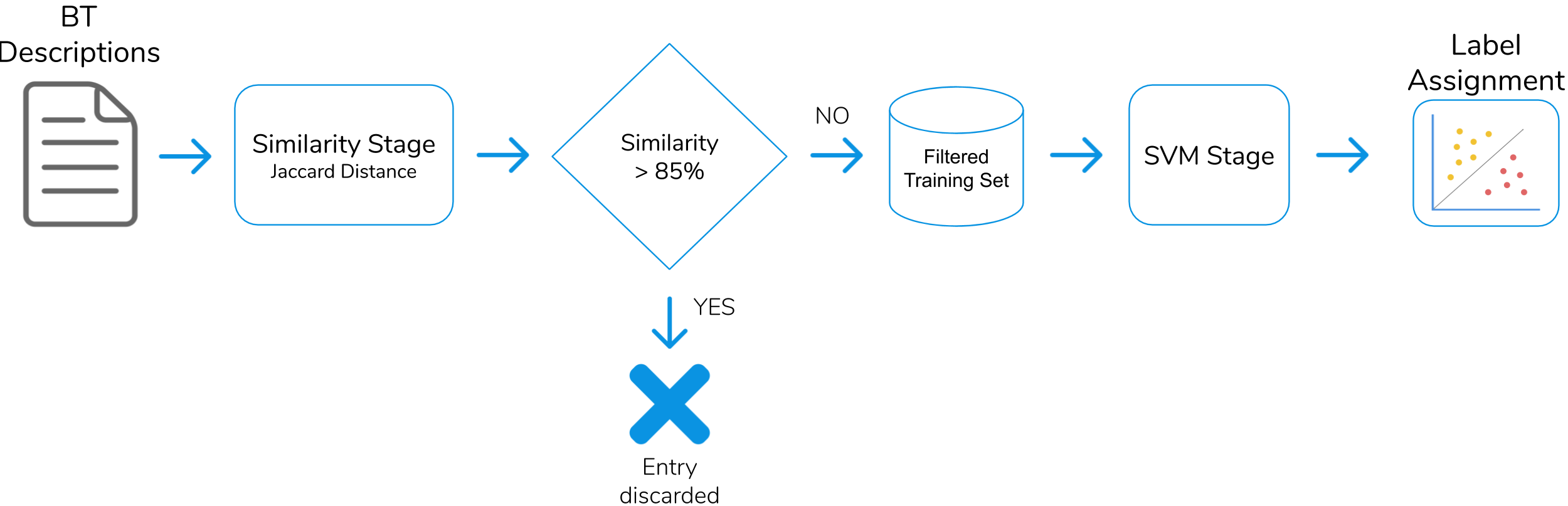}
\caption{\label{similarity}Flow diagram of the system with Jaccard similarity detector.}
\end{figure*}

Figure \ref{similarity} illustrates the architecture of the system including the Jaccard similarity detector. The {\sc svm} classifier is explained in Section \ref{svm}.

\subsection{Machine learning analysis}

In this section we explain the knowledge-based linguistic extraction as well as the feature selection. 

\subsubsection{Linguistic knowledge extraction}
\label{linguistic}

\begin{figure*}[!htbp]
\centering
\includegraphics[width=0.6\textwidth]{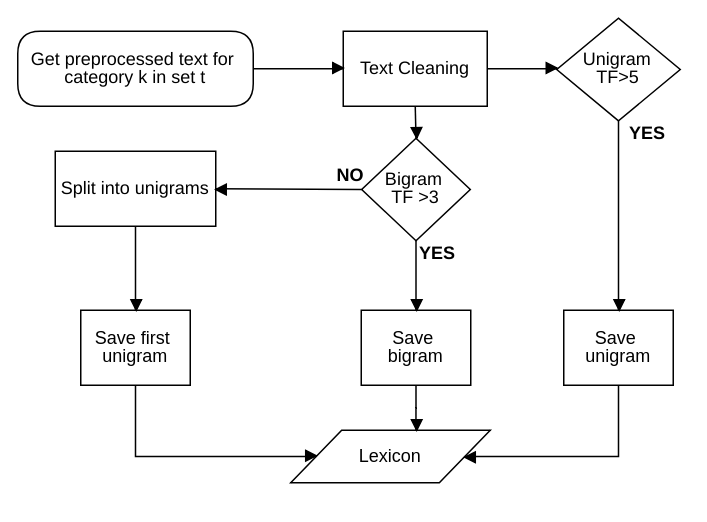}
\caption{\label{lexicaUML}{\sc uml} diagram of the lexicon generation procedure.}
\end{figure*}

In this step we create lexica whose entries are related to the categories of the classification problem. Figure~\ref{lexicaUML} represents the lexicon generation procedure. 

First, starting from the preprocessed {\sc bt} descriptions in the training set, which are labelled according to the classification categories, all non-alphabetic characters such as numbers, punctuation marks and symbols are cleared. Next, useful final elements for the lexica are extracted. These are the unigrams that appear at least five times in the text corpora for each category (all others are excluded) and the bigrams that are present at least three times in the corpora. Single-character alphabetic elements are also discarded. The final result is a set of lexica with unigrams and bigrams and their corresponding categories. 

For example, let us suppose that the training set only has the following entries for a given category:
\begin{enumerate}
\item {\em Compra en Pescados Diego, S.L.} (`Purchase at Pescados Diego, S.L.')
\item {\em Compra en supermercado Elvira Madrid 28} (`Purchase at Elvira supermarket Madrid 28')
\item {\em Compra en amazon.es} (`Purchase in amazon.es')
\item {\em Compra en supermercado Carrefour Enero 2018} (`Purchase at Carrefour supermarket January 2018')
\item {\em Compra en amazon.es Febrero 2018} (`Purchase in amazon.es February 2018')
\item {\em Compra en Amazon} (``Purchase in Amazon')
\item {\em Pago en supermercado Elvira Alicante} (`Payment at Elvira Alicante supermarket')
\item {\em Pago en supermercado El Corte Ingl\'es Vigo} (`Payment at El Corte Inglés Vigo supermarket')
\item {\em Compra en supermercado Carrefour Febrero 2018} (`Purchase at Carrefour supermarket February 2018')
\item {\em Compra en supermercado amazon.es} (`Purchase in amazon.es supermarket')
\end{enumerate}

The resulting lexicon would only contain the words {\em `compra'} and {\em `supermercado'} and the bigram {\em `compra supermercado'} followed by the categories. 

\subsubsection{Feature selection and weight calculation}
\label{featuresSVM}

The system uses a standard {\sc svm} algorithm for modelling and prediction. Short texts are encoded according to the vector space model in~\cite{yin2014towards}. The smallest data unit in the model corresponds to a feature. A text $ T $ may be seen as an $n$-dimensional vector in the vector space, as follows:

\begin{equation}
T = ((t_1,w_1),(t_2,w_2),...,(t_n,w_n))
\end{equation}

\noindent where $ t $ is the value of a feature  of text $ T $ and $ w $ its weight. The greater the $ w $, the more information the feature contains in that case \cite{yin2013botnet}.

Many different types of features are possible, such as Boolean, word frequency (number of times a word appears in the text) and {\sc tf-idf}. Note that classification results depend greatly on feature selection \cite{veeraswamy2011survey,tong2007research}. An efficient feature selection method not only reduces the dimension of the feature space but also avoids useless features. The features in our system are the following:

\begin{enumerate}
\item {Lexicon data}. These features count the words in the {\sc bt} descriptions that appear in the lexica for each existing category.

\item {Amount}. The range of the {\sc bt} amount field, since ranges are more significant for our application than exact values. Specifically, we consider non-overlapping intervals limited by  
$ 20 $, $ 60$, $ 200 $, $ 800 $, $ 1500 $ and $ 3000 $ euros.

\item {Sign of the amount}. This feature indicates if the {\sc bt} is an income (positive) or an expense (negative).

\item {Date}. The information in the date field of each {\sc bt}. Again, we use ranges. This is because some events occur on specific days of the month (e.g. salary at the end), whereas other events (e.g. purchases) may happen anytime during the month. The selected ranges were the last five, ten, twenty and twenty-five days of the month.

\item {Word $n$-grams}. $N$-gram representation is language-independent. It transforms documents into high-dimensional feature vectors where each feature corresponds to a contiguous sub-string. Formally, an $n$-gram consists of $n$ adjacent items from alphabet $A$. Items can be phonemes, syllables, letters, words or base pairs depending on the application. Hence, the number of different $n$-grams in a text is $ |A|^n $ at most. The dimension of an $n$-gram feature sub-vector may therefore be very high even for moderate values of $n$. However, since not all $n$-grams are present in a document, the dimension is substantially reduced. During the formation of an $n$-gram feature sub-vector, all upper-case characters are converted into lower-case characters and punctuation marks are converted to spaces. Sub-vectors are then normalized. The optimal $n$ depends on the text corpora.

We explain feature sub-vectors with an example that computes the $n$-grams from one to four words for the {\sc bt} description {\em `Operación tarjeta débito Amazon'} (`Amazon debit card transaction'). The resulting vector consists of the following components: {\em `operación'} (`transaction'), {\em `tarjeta'} (`card'), {\em `débito'} (`debit'), {\em amazon}; {\em `operación tarjeta'} (`card transaction'), {\em `tarjeta débito'} (`debit card'), {\em `débito amazon'} (`amazon debit'); {\em `operación tarjeta débito'} (`debit card transaction'), {\em `tarjeta débito amazon'} (`amazon debit card'); {\em `operación tarjeta débito amazon'} (`amazon debit card transaction').

\item {Character $n$-grams}. 
Character $n$-grams have been proven useful for a variety of {\sc ml} problems, such as language detection. Simple models based on them have outperformed convolutional and recursive deep neural networks ({\sc cnn}s and {\sc rnn}s) \cite{Medvedeva2017, Malmasi17, Kulmizev17}. 

We illustrate them with an example that computes the trigram, four-gram and five-gram character sub-vectors for the sentence {\em `Operación tarjeta débito Amazon'} (note that spaces are also taken into account when computing character $n$-grams): ({\em ope, per, era, rac, aci, ció, ión, ón , n t,  ta, tar, arj, rje, jet, eta, ta , a d,  dé, déb, ébi, bit, ito, to , o a,  am, ama, maz, azo, zon}; {\em oper, pera, erac, raci, ació, ción, ión , ón t, n ta,  tar, tarj, arje, rjet, jeta, eta , ta d, a dé,  déb, débi, ébit, bito, ito , to a, o am,  ama, amaz, mazo, azon}; {\em opera, perac, eraci, ració, ación, ción , ión t, ón ta, n tar,  tarj, tarje, arjet, rjeta, jeta , eta d, ta dé, a déb,  débi, débit, ébito, bito , ito a, to am, o ama,  amaz, amazo, mazon}).

They have been applied in scenarios with misspelling errors \cite{cavnar1995using,huffman1995acquaintance}. Character $n$-grams may also capture other effects of language usage, such as re-named entities and abbreviations, e.g. `maths' instead of `mathematics'. In our case, they are justified by the many shortened words in {\sc bt} descriptions.

\end{enumerate}

\subsubsection{{\sc svm} classifier}
\label{svm}

We decompose the overall problem into pairwise two-class problems, following a one-versus-one approach. Therefore, $k(k-1)/2$ {\sc svm} classification models are necessary for $k$ text classes. The category is decided by majority voting. 

\section{Experimental results}
\label{results}

All experiments were performed on a computer with the following specifications:

\begin{enumerate}
    \item Operating System: Ubuntu 18.04 LTS 64 bits
    \item Processor: Intel@Core i5 3470 CPU 3.2Ghz x 4 
    \item RAM: 15.4 Gb
    \item Disk: 1.9 Tb
\end{enumerate}

\subsection{Dataset}
\label{sec-dataset}
The dataset comprises 30,844 {\sc bt} descriptions from customer accounts of major Spanish banks, written mostly in Spanish and issued between August 2017 and February 2018. They were collected during the CatCoin project with the collaboration of CoinScrap Finance S.L., Spain, using the {\em CoinScrap} platform. The entries of the dataset have the following attributes:

\begin{enumerate}
\item ID: a unique numeric identifier.
\item Description: the {\sc bt} short-text description.
\item Amount: the amount in euros of the {\sc bt}, either positive (income) or negative (expense).
\item Date: the date when the {\sc bt} occurred.
\end{enumerate} 

Every entry has an extra field with the category label that determines the classification goal. The dataset may be requested to the authors by e-mail. Table~\ref{categoriesDist} shows the numerical distributions of the fifteen categories in the dataset. Table~\ref{entriesExample} shows some examples of dataset entries.
\begin{table*}[!htbp]
\centering
\caption{\label{categoriesDist}Distribution of categories in the labelled dataset.}
\begin{tabular}{lc}\hline
\multicolumn{1}{c}{Category} & \multicolumn{1}{c}{Instances} \\ \hline
Bank & 4,835 \\
Means of transport & 3,479 \\
Shopping & 11,061 \\
Household expenses & 1,158 \\
Taxes and charges & 489 \\
Off-cycle income & 89 \\
Payroll & 248 \\
Leisure & 2,362 \\
Health, sport and education & 867 \\
Insurances & 883 \\
Social security, grants and pensions & 67 \\
Transfers & 2,086 \\
Business and professional expenses & 197 \\
Rentals & 116 \\
Others & 2,907 \\\hline
\multicolumn{1}{l}{Total} & 30,844 \\
\hline
\end{tabular}
\end{table*}

\subsection{Evaluation metrics}

Due to the issues of accuracy with class asymmetries \cite{SokolovaEtAl09,RossiEtAl16}, we employed precision,  recall and $F$ metrics using a macro-average approach. 

\begin{table*}[!htbp]
\centering
\caption{\label{entriesExample}Examples of entries in the labelled dataset.}
\begin{tabular}{lllcl}\hline
\multicolumn{1}{c}{ID} & \multicolumn{1}{c}{Description} &
\multicolumn{1}{c}{Category} & \multicolumn{1}{c}{Amount} & 
\multicolumn{1}{c}{Date} \\ \hline
\begin{tabular}[c]{@{}l@{}}59da944c5858\\ aa32256f883a\end{tabular} & \begin{tabular}[c]{@{}l@{}}Recibo ORANGE ESPAGNE S.A.U\\`{\em ORANGE ESPAGNE S.A.U bill}'\end{tabular} & Household expenses &-42,29 € & \begin{tabular}[c]{@{}l@{}}Thu Sep 28 2017\\ 02:00:00\end{tabular} \\\hline
\begin{tabular}[c]{@{}l@{}}59da944c5858\\ aa32256f882a\end{tabular}	& \begin{tabular}[c]{@{}l@{}}Traspaso recibido Cuenta Nómina\\`{\em Transfer received Payroll Account}'\end{tabular} & Bank & 100,00 € & \begin{tabular}[c]{@{}l@{}}Thu Oct 05 2017\\ 02:00:00\end{tabular}\\\hline
\begin{tabular}[c]{@{}l@{}}5a046cf2d9f7\\ 0921c74182d9\end{tabular}	& www.just-eat.es & Leisure & -39,60	€ &\begin{tabular}[c]{@{}l@{}}Thu Nov 09 2017\\ 01:00:00\end{tabular}\\\hline
\begin{tabular}[c]{@{}l@{}}5a69353d323c\\ a817506a2bdd\end{tabular} & \begin{tabular}[c]{@{}l@{}}RECIBO ASOC DE CONSUMIDORES\\ EN ACCION-FACUA\\ `{\em Association of Consumers in}\\ {\em Action-FACUA bill}'\end{tabular} & Health, sport and education & -63,00	€ & \begin{tabular}[c]{@{}l@{}}Tue Jan 09 2018\\ 01:00:00\end{tabular}\\\hline
\begin{tabular}[c]{@{}l@{}}5a8d5a8d1a33\\ 590273326bed\end{tabular} & \begin{tabular}[c]{@{}l@{}}TRANSFERENCIA A FAVOR DE PN\\CONCEPTO Alquiler Febrero 2017 + Luz\\`{\em Transfer in favour of PN CONCEPT} \\{\em February 2017 rent + electricity}'\end{tabular} & Rentals & -377,75 € & \begin{tabular}[c]{@{}l@{}}Wed Feb 07 2018\\ 01:00:00\end{tabular} \\\hline
\end{tabular}
\end{table*}

Macro-averaged results were computed as indicated by~\cite{tsoumakas2009mining}. Consider a binary evaluation metric $B(t_p, t_n, f_p, f_n)$ that is calculated based on the number of true positives ($t_p$), true negatives ($t_n$), false positives ($f_p$) and false negatives ($f_n$). Let $t_{p_\lambda}$, $f_{p_\lambda}$, $t_{n_\lambda}$ and $ f_{n_\lambda}$ be the amounts of true positives, false positives, true negatives and false negatives, respectively, after binary evaluation for label $\lambda$. The macro-average evaluation metric is calculated as follows:

\begin{equation}
\small
\displaystyle B_{macro} = \frac{1}{q} \sum_{\lambda=1}^k B(t_{p_\lambda}, f_{p_\lambda}, t_{n_\lambda}, f_{n_\lambda})
\label{eq-1}
\end{equation}

Macro-averaging weights all classes equally, whereas micro-averaging weights all document classification decisions equally. Since $F$ ignores true negatives and its magnitude is mostly determined by the number of true positives, large classes dominate over small classes in micro-averaging \cite{manning2008introduction}. For this reason we preferred the macro-average approach.

To calculate precision, recall and $F$ rates we first computed each of these measures separately for each category using expressions~\eqref{eq-2}-\eqref{eq-4}:

\begin{equation}
\small
\displaystyle \mbox{\em Precision}_{\mbox{\footnotesize micro}_{q}} = \frac{t_{p_q}}{t_{p_q}+f_{p_q}}
\label{eq-2}
\end{equation}

\begin{equation}
\small
\displaystyle \mbox{\em Recall}_{\mbox{\footnotesize micro}_{q}} = \frac{t_{p_q}}{t_{p_q}+f_{n_q}}
\label{eq-3}
\end{equation}

\begin{equation}
\small
\displaystyle \mbox{\em F}_{\mbox{\footnotesize micro}_{q}} = \frac{2(\mbox{\em Precision}_{\mbox{\footnotesize micro}_{q}}*\mbox{\em Recall}_{\mbox{\footnotesize micro}_{q}})}{(\mbox{\em Precision}_{\mbox{\footnotesize micro}_{q}}+\mbox{\em Recall}_{\mbox{\footnotesize micro}_{q}})}
\label{eq-4}
\end{equation}

These metrics were then averaged by category using expression ~\eqref{eq-1} to produce the macro-averaged metrics.

\subsection{Numerical results}

We performed cross-validation in different dataset splits of training and testing subsets (in all cases the first and second percentages correspond to training and testing subset sizes, respectively): 30\%-70\%, 40\%-60\%, 60\%-40\% and 70\%-30\%. The purpose was to  check the robustness of our system when fewer training data were available. 

\begin{table*}[!htbp]
\centering
\caption{\label{lexicaTable}Average word distribution in the lexica for the different training-testing splits before applying the similarity filter.}
\begin{tabular}{lcccc}\hline
\multicolumn{1}{c}{Lexicon} & \multicolumn{1}{c}{{30\%-70\%}} & \multicolumn{1}{c}{{40\%-60\%}} & \multicolumn{1}{c}{{60\%-40\%}} & \multicolumn{1}{c}{{70\%-30\%}}\\ \hline
Bank & 368.6 & 421.8 & 486.2 & 513.6 \\
Means of transport & 515 & 602.6 & 742.4 & 782.6 \\
Shopping & 1,551.2 & 1,830.2 & 2,295.2 & 2,501.4 \\
Household expenses & 191.4 & 226.8 & 280.4 & 307.6 \\
Taxes and charges & 101.8 & 128.4 & 159.8 & 169.2 \\
Off-cycle income & 13.2 & 18.2 & 24.2 & 113.2 \\
Payroll & 65 & 80.4 & 105 & 27.6 \\
Leisure & 479.6 & 569.8 & 706.4 & 765.2 \\
Health, sport and education & 262 & 310.2 & 393.2 & 437.2 \\
Insurances & 154 & 181.2 & 244.6 & 259.4 \\
Social security, grants and pensions & 15.2 & 19.2 & 23.4 & 24.2 \\
Transfers & 516.6 & 639.6 & 849.6 & 933 \\
Business and professional expenses & 61.8 & 80.4 & 107.8 & 119.6 \\
Rentals & 54.8 & 58.6 & 79.2 & 90 \\
Others & 160.2 & 190.6 & 226.4 & 241.8\\
\hline
\end{tabular}
\end{table*}

In each experiment we extracted the lexica of the set as explained in Section~\ref{linguistic}. Table~\ref{lexicaTable} shows the distributions of words in the lexica for all categories before applying the similarity detector. We added features incrementally to the model to assess their significance. Therefore, first we only used word $n$-grams and lexica, then we added {\sc bt} amount and date, and finally character $n$-grams features. 

Given the target sector (finance), precision may be more important than recall. This is because banking campaigns prefer to obtain less positives for key categories. By doing so, they maximize the probability that customers will be receptive to  personalized products.

We compared our system with three competitor approaches, All-In-1 \cite{plank17} and two variants of the method by IITP (Indian Institute of Technology Patna) \cite{IITP17}. These approaches  analyzed customer feedback to manufacturers, which also consisted of short texts, although with more elaborate sentences than {\sc bt} descriptions. Note that no other researchers have considered {\sc bt} to date. For the sake of fairness, we applied the Jaccard distance detector stage to the competitors as well.

The All-In-1 approach in \cite{plank17} is based on a classic {\sc svm} classifier that takes character $n$-grams and monolingual word embeddings as input. Logically we only used the monolingual version. 

The two IITP variants  \cite{IITP17} are based on {\sc cnn}s. The second variant combines a {\sc cnn} with an {\sc rnn}. Specifically, a convolutional feature extractor is applied to the input, a recurrent network is applied to the {\sc cnn} output, an optional fully connected layer is applied to the {\sc rnn} output, and finally a softmax layer delivers the result. 

Tables~\ref{evaluationTime3} and~\ref{evaluationTimeOtros} show average elapsed training and testing times for our system and its competitors for the different splits and selections of features, obtained with cross-validation (five different dataset samplings in each experiment). For our system, the values of $n$ in character  and word $n$-grams were adjusted to 3-5 and 1-4  respectively.

Note that, even though testing times were comparable,   the training times of the competitors were significantly higher. This is due to their greater computational complexity. Specifically,  the {\sc svm} classifier of All-In-1 uses word embeddings  and the two variants of IITP are based on {\sc cnn}s.

Table \ref{reduction_training} shows the average training set reductions that the similarity detector achieved for the different splits. Note that they exceeded 56\% in all cases.

\begin{table*}[!htbp]
\centering
\caption{\label{evaluationTime3}Elapsed training and testing times of our system for different dataset splits.}
\begin{tabular}{clclc}
\hline
\multicolumn{1}{c}{Split} & \multicolumn{1}{c}{Set} & \multicolumn{1}{c}{Instances} & \multicolumn{1}{c}{Features} & \multicolumn{1}{c}{Computing time (s)} \\ \hline
\multirow{10}{*}{30\%-70\%} & \multirow{5}{*}{Train} & \multirow{5}{*}{4,031} & Word $n$-grams & 2.20 $\pm$ 0.40\\
& & & Word $n$-grams + char $n$-grams & 5.00 $\pm$ 1.55\\
& & & Word $n$-grams + lexica & 2.20 $\pm$ 0.40\\
& & & Word $n$-grams + lexica + amount + date & 2.20 $\pm$ 0.40\\
& & & Word $n$-grams + lexica + amount + date + char $n$-grams & 6.20 $\pm$ 1.47\\
\cline{2-5}
 & \multirow{5}{*}{Test} & \multirow{5}{*}{21,590} & Word $n$-grams & 8.20 $\pm$ 0.40\\
& & & Word $n$-grams + char $n$-grams & 15.40 $\pm$ 0.49\\
& & & Word $n$-grams + lexica & 10.00 $\pm$ 0.00\\
& & & Word $n$-grams + lexica + amount + date & 10.40 $\pm$ 0.49\\
& & & Word $n$-grams + lexica + amount + date + char $n$-grams & 18.60 $\pm$ 2.15\\
\hline
\multirow{10}{*}{40\%-60\%} & \multirow{5}{*}{Train} & \multirow{5}{*}{4,849.40} & Word $n$-grams & 2.00 $\pm$ 0.00\\
& & & Word $n$-grams + char $n$-grams & 5.80 $\pm$ 0.75\\
& & & Word $n$-grams + lexica & 3.00 $\pm$ 0.00\\
& & & Word $n$-grams + lexica + amount + date & 3.20 $\pm$ 0.40\\
& & & Word $n$-grams + lexica + amount + date + char $n$-grams & 6.00 $\pm$ 0.00\\
\cline{2-5}
 & \multirow{5}{*}{Test} & \multirow{5}{*}{18,506} & Word $n$-grams & 7.00 $\pm$ 0.00\\
& & & Word $n$-grams + char $n$-grams & 14.00 $\pm$ 0.00\\
& & & Word $n$-grams + lexica & 8.40 $\pm$ 0.49\\
& & & Word $n$-grams + lexica + amount + date & 8.40 $\pm$ 0.49\\
& & & Word $n$-grams + lexica + amount + date + char $n$-grams & 15.20 $\pm$ 0.40\\
\hline
\multirow{10}{*}{60\%-40\%} & \multirow{5}{*}{Train} & \multirow{5}{*}{6,209.20} & Word $n$-grams & 3.00 $\pm$ 0.00\\
& & & Word $n$-grams + char $n$-grams & 9.40 $\pm$ 1.50\\
& & & Word $n$-grams + lexica & 4.20 $\pm$ 0.40\\
& & & Word $n$-grams + lexica + amount + date &  4.00 $\pm$ 0.00\\
& & & Word $n$-grams + lexica + amount + date + char $n$-grams & 9.00 $\pm$ 0.63\\
\cline{2-5}
 & \multirow{5}{*}{Test} & \multirow{5}{*}{12,338} & Word $n$-grams & 5.00 $\pm$ 0.00\\
& & & Word $n$-grams + char $n$-grams & 11.40 $\pm$ 0.80\\
& & & Word $n$-grams + lexica & 6.20 $\pm$ 0.40\\
& & & Word $n$-grams + lexica + amount + date & 6.00 $\pm$ 0.00\\
& & & Word $n$-grams + lexica + amount + date + char $n$-grams & 12.00 $\pm$ 0.00\\
\hline
\multirow{10}{*}{70\%-30\%} & \multirow{5}{*}{Train} & \multirow{5}{*}{6,780.20} & Word $n$-grams & 4.00 $\pm$ 0.00\\
& & & Word $n$-grams + char $n$-grams & 11.20 $\pm$ 0.98\\
& & & Word $n$-grams + lexica & 4.20 $\pm$ 0.40\\
& & & Word $n$-grams + lexica + amount + date & 4.60 $\pm$ 0.80\\
& & & Word $n$-grams + lexica + amount + date + char $n$-grams & 11.20 $\pm$ 0.40\\
\cline{2-5}
 & \multirow{5}{*}{Test} & \multirow{5}{*}{9,253} & Word $n$-grams & 4.00 $\pm$ 0.00\\
& & & Word $n$-grams + char $n$-grams & 9.60 $\pm$ 0.49\\
& & & Word $n$-grams + lexica & 5.20 $\pm$ 0.40\\
& & & Word $n$-grams + lexica + amount + date & 5.20 $\pm$ 0.40\\
& & & Word $n$-grams + lexica + amount + date + char $n$-grams & 10.60 $\pm$ 0.49\\
\hline
\end{tabular}
\end{table*}

\begin{table*}[!htbp]
\centering
\caption{\label{evaluationTimeOtros}Elapsed training and testing times of the competitor systems for different dataset splits.}
\small
\begin{tabular}{clclc}
\hline
\multicolumn{1}{c}{Split} & \multicolumn{1}{c}{Set} & \multicolumn{1}{c}{Instances} & \multicolumn{1}{c}{Features} & \multicolumn{1}{c}{Computing time (s)} \\ \hline
\multirow{12}{*}{30\%-70\%} & \multirow{6}{*}{Train} & \multirow{6}{*}{4,031} & Word $n$-grams & 2.20 $\pm$ 0.40\\
& & & Word $n$-grams + char $n$-grams & 5.00 $\pm$ 1.55\\
\cline{4-5}

& & & All-In-1 Word $n$-grams & 1.83 $\pm$ 0.08\\
& & & All-In-1 Word $n$-grams + char $n$-grams & 30.04 $\pm$ 1.66\\
& & & IITP-{\sc cnn} & 976.10 $\pm$ 33.34\\
& & & IITP-{\sc cnn+rnn} &  1153.03 $\pm$ 59.08\\
\cline{2-5}
 & \multirow{6}{*}{Test} & \multirow{6}{*}{21,590} & Word $n$-grams & 8.20 $\pm$ 0.40\\
& & & Word $n$-grams + char $n$-grams & 15.40 $\pm$ 0.49\\
\cline{4-5}

& & & All-In-1 Word $n$-grams &  1.56 $\pm$ 0.02\\
& & & All-In-1 Word $n$-grams + char $n$-grams &  5.53 $\pm$ 0.16\\
& & & IITP-{\sc cnn} &  4.47 $\pm$ 0.37\\
& & & IITP-{\sc cnn+rnn} & 5.99 $\pm$ 0.62\\

\hline
\multirow{12}{*}{40\%-60\%} & \multirow{6}{*}{Train} & \multirow{6}{*}{4,849.40} & Word $n$-grams & 2.00 $\pm$ 0.00\\
& & & Word $n$-grams + char $n$-grams & 5.80 $\pm$ 0.75\\
\cline{4-5}

& & & All-In-1 Word $n$-grams &  2.28 $\pm$ 0.15\\
& & & All-In-1 Word $n$-grams + char $n$-grams & 38.68 $\pm$ 1.70\\
& & & IITP-{\sc cnn} & 1224.89 $\pm$ 71.48\\
& & & IITP-{\sc cnn+rnn} & 1698.12 $\pm$ 108.46\\

\cline{2-5}
 & \multirow{6}{*}{Test} & \multirow{6}{*}{18,506} & Word $n$-grams & 7.00 $\pm$ 0.00\\
& & & Word $n$-grams + char $n$-grams & 14.00 $\pm$ 0.00\\
\cline{4-5}

& & & All-In-1 Word $n$-grams &  1.33 $\pm$ 0.04\\
& & & All-In-1 Word $n$-grams + char $n$-grams & 5.01 $\pm$ 0.12\\
& & & IITP-{\sc cnn} & 3.85 $\pm$ 0.28\\
& & & IITP-{\sc cnn+rnn} & 5.80 $\pm$ 0.19\\

\hline
\multirow{12}{*}{60\%-40\%} & \multirow{6}{*}{Train} & \multirow{6}{*}{6,209.20} & Word $n$-grams & 3.00 $\pm$ 0.00\\
& & & Word $n$-grams + char $n$-grams & 9.40 $\pm$ 1.50\\
\cline{4-5}

& & & All-In-1 Word $n$-grams & 2.93 $\pm$ 0.08\\
& & & All-In-1 Word $n$-grams + char $n$-grams & 61.95 $\pm$ 3.68\\
& & & IITP-{\sc cnn} & 2025.20 $\pm$ 22.99\\
& & & IITP-{\sc cnn+rnn} &  2759.47 $\pm$ 66.43\\

\cline{2-5}
 & \multirow{6}{*}{Test} & \multirow{6}{*}{12,338} & Word $n$-grams & 5.00 $\pm$ 0.00\\
& & & Word $n$-grams + char $n$-grams & 11.40 $\pm$ 0.80\\
\cline{4-5}

& & & All-In-1 Word $n$-grams & 0.91 $\pm$ 0.02\\
& & & All-In-1 Word $n$-grams + char $n$-grams & 3.91 $\pm$ 0.16\\
& & & IITP-{\sc cnn} & 2.90 $\pm$ 0.09\\
& & & IITP-{\sc cnn+rnn} & 4.45 $\pm$ 0.13\\

\hline
\multirow{12}{*}{70\%-30\%} & \multirow{6}{*}{Train} & \multirow{6}{*}{6,780.20} & Word $n$-grams & 4.00 $\pm$ 0.00\\
& & & Word $n$-grams + char $n$-grams & 11.20 $\pm$ 0.98\\
\cline{4-5}

& & & All-In-1 Word $n$-grams & 3.21 $\pm$ 0.09\\
& & & All-In-1 Word $n$-grams + char $n$-grams & 69.35 $\pm$ 1.75\\
& & & IITP-{\sc cnn} & 2327.73 $\pm$ 38.76\\
& & & IITP-{\sc cnn+rnn} & 3072.74 $\pm$ 327.81\\
\cline{2-5}

 & \multirow{6}{*}{Test} & \multirow{6}{*}{9,253} & Word $n$-grams & 4.00 $\pm$ 0.00\\
& & & Word $n$-grams + char $n$-grams & 9.60 $\pm$ 0.49\\
\cline{4-5}

& & & All-In-1 Word $n$-grams & 0.69 $\pm$ 0.02\\
& & & All-In-1 Word $n$-grams + char $n$-grams & 3.30 $\pm$ 0.10\\
& & & IITP-{\sc cnn} & 2.37 $\pm$ 0.06\\
& & & IITP-{\sc cnn+rnn} & 3.41 $\pm$ 0.12\\
\hline
\end{tabular}
\end{table*}

\begin{table*}[!htbp]
\centering
\caption{\label{reduction_training}Training sample reduction for different dataset splits.}
\begin{tabular}{ccc}
\hline
Split & Instances & Instances after similarity detection\\ \hline
\multirow{1}{*}{30\%-70\%} & 9,254 & 4,031\\
\multirow{1}{*}{40\%-60\%} & 12,338 & 4,849\\
\multirow{1}{*}{60\%-40\%} & 18,506 & 6,209\\
\multirow{1}{*}{70\%-30\%} & 21,591 & 6,780\\
\hline
\end{tabular}
\end{table*}

\subsubsection{30\%-70\% split}
In each experiment the training dataset had in average 4,031 entries (after similarity detection) and the testing dataset had 21,590 entries.

Tables~\ref{evaluation3070Basico} and~\ref{evaluation3070Nuevo} show the results of {\sc bt} classification. Note that we did not modify the design or the implementation of the selected competitors. Thus, for a fair comparison, only word and character $n$-grams features were enabled in Table~\ref{evaluation3070Basico}. Our system outperformed the competitors in terms of precision and All-In-1 was the best option in recall and $F$-measure.

\begin{table*}[!htbp]
\centering
\caption{\label{evaluation3070Basico}Average evaluation metrics for the basic combinations of features, 30\%-70\% split.}
\small
\begin{tabular}{ccp{8.5cm}ccc}
\hline
\multicolumn{1}{c}{Train} & \multicolumn{1}{c}{Test} & \multicolumn{1}{c}{Features} & \multicolumn{1}{c}{$\mbox{P}_{\mbox{\tiny \bf macro}}$} & \multicolumn{1}{c}{ $\mbox{R}_{\mbox{\tiny \bf macro}}$} & \multicolumn{1}{c}{$\mbox{F}_{\mbox{\tiny \bf macro}}$}\\ \hline
\multirow{6}{*}{30\%} & \multirow{6}{*}{70\%} 
& Proposed system word $n$-grams & 68.19\% & 25.70\% & 37.32\% \\
& & Proposed system word $n$-grams + char $n$-grams & \bf 93.36\% & 76.30\% & 83.97\% \\\cline{3-6}
& & All-In-1 word $n$-grams & 90.87\% & 85.22\% & 87.67\% \\
& & All-In-1 word $n$-grams + char $n$-grams & 90.75\% & \bf 86.87\% & \bf 88.57\% \\
& & IITP-{\sc cnn}  & 87.83\% & 81.02\% & 83.78\%\\
& & IITP-{\sc cnn}+{\sc rnn} & 88.27\% & 75.05\% & 80.32\%\\
\hline
\end{tabular}
\end{table*}

In Table~\ref{evaluation3070Nuevo} we observe that, after activating the lexicon feature, the precision, recall and $F$ of our system increased by about 15\%, 38\% and 35\%, respectively, so the lexicon feature was crucial. Another key result is that meta-information features yielded a precision increase of 8\% in our system. After activating all features, precision, recall and $F$ further improved by around 3\%, 19\% and 13\%, respectively.

\begin{table*}[!htbp]
\centering
\caption{\label{evaluation3070Nuevo}Average evaluation metrics of the proposed system for all combinations of features, 30\%-70\% split.}
\small
\begin{tabular}{ccp{8.5cm}ccc}
\hline
\multicolumn{1}{c}{Train} & \multicolumn{1}{c}{Test} & \multicolumn{1}{c}{Features} & \multicolumn{1}{c}{$\mbox{P}_{\mbox{\tiny \bf macro}}$} & \multicolumn{1}{c}{ $\mbox{R}_{\mbox{\tiny \bf macro}}$} & \multicolumn{1}{c}{$\mbox{F}_{\mbox{\tiny \bf macro}}$}\\ \hline
\multirow{4}{*}{30\%} & \multirow{4}{*}{70\%}
& Word $n$-grams & 68.19\% & 25.70\% & 37.32\% \\
& & Word $n$-grams + lexica & 82.90\% & 63.51\% & 71.90\%\\
& & Word $n$-grams + lexica + amount + date & 91.31\% & 64.92\% & 75.87\%\\
& & Word $n$-grams + lexica + amount + date + char $n$-grams  & \bf 94.59\% & \bf 84.15\% & \bf 89.06\% \\
\hline
\end{tabular}
\end{table*}

Our system attained the best precision, but only attained better $F$ than All-In-1 if all features are activated. On the other hand, All-In-1 was better in recall but the difference with our system in that regard was only about 3\%. Note, however, that regardless of the fact that precision is more important in our scenario, our system is simpler than its competitors (based, depending on the case, on {\sc cnn}, {\sc cnn}+{\sc rnn} or a {\sc svm} with word embeddings), especially in terms of training time, as shown in tables \ref{evaluationTime3} and \ref{evaluationTimeOtros}. 

\subsubsection{40\%-60\% split}
In this case, in each experiment the training dataset had in average 4,849 annotated entries (after similarity detection) and the testing dataset had 18,506 entries.

Tables~\ref{evaluation4060Basico} and~\ref{evaluation4060Nuevo} show the results. In this case our improvement in precision with the basic features was about 2\% compared to the competitors. The precision, recall and $F$ of our system increased by about 17\%, 37\% and 34\%, respectively, after activating the lexica feature. By adding meta-information, the improvements were  7\%, 3\% and 5\%, respectively. Total precision, recall and $F$ improved compared to the baseline (word $n$-grams) by about 2\%, 18\% and 12\%, respectively, after activating all the features.

\begin{table*}[!htbp]
\centering
\caption{\label{evaluation4060Basico}Average evaluation metrics for the basic combinations of features, 40\%-60\% split.}
\small
\begin{tabular}{ccp{8.3cm}ccc}
\hline
\multicolumn{1}{c}{Train} & \multicolumn{1}{c}{Test} & \multicolumn{1}{c}{Features} & \multicolumn{1}{c}{$\mbox{P}_{\mbox{\tiny \bf macro}}$} & \multicolumn{1}{c}{$\mbox{R}_{\mbox{\tiny \bf macro}}$} & \multicolumn{1}{c}{$\mbox{F}_{\mbox{\tiny \bf macro}}$}\\ \hline
\multirow{6}{*}{40\%} & \multirow{6}{*}{60\%} 
& Proposed system word $n$-grams & 68.25\% & 27.36\% & 39.05\% \\
& & Proposed system word $n$-grams + char $n$-grams & \bf 93.56\% & 78.72\% & 85.50\%\\\cline{3-6}
& & All-In-1 word $n$-grams & 91.69\% & 87.12\% & 89.05\% \\
& & All-In-1 word $n$-grams + char $n$-grams & 90.99\% & 88.11\% & \bf 89.27\% \\
& & IITP-{\sc cnn}  & 90.27\% & \bf 84.06\% & 86.70\%\\
& & IITP-{\sc cnn}+{\sc rnn} & 88.84\% & 77.93\% & 82.15\%\\
\hline
\end{tabular}
\end{table*}

\begin{table*}[!htbp]
\centering
\caption{\label{evaluation4060Nuevo}Average evaluation metrics of the proposed system for all combinations of features, 40\%-60\% split.}
\small
\begin{tabular}{ccp{8.5cm}ccc}
\hline
\multicolumn{1}{c}{Train} & \multicolumn{1}{c}{Test} & \multicolumn{1}{c}{Features} & \multicolumn{1}{c}{$\mbox{P}_{\mbox{\tiny \bf macro}}$} & \multicolumn{1}{c}{ $\mbox{R}_{\mbox{\tiny \bf macro}}$} & \multicolumn{1}{c}{$\mbox{F}_{\mbox{\tiny \bf macro}}$}\\ \hline
\multirow{4}{*}{40\%} & \multirow{4}{*}{60\%}
& Word $n$-grams & 68.25\% & 27.36\% & 39.05\% \\
& & Word $n$-grams + lexica & 85.54\% & 64.50\% & 73.53\%\\
& & Word $n$-grams + lexica + amount + date & 92.94\% & 67.30\% & 78.05\%\\
& & Word $n$-grams + lexica + amount + date + char $n$-grams  & \bf 95.43\% & \bf 85.43\% & \bf 90.15\% \\
\hline
\end{tabular}
\end{table*}

We again attained the best precision performance, outperforming the best competitor by almost 4\% when all features are activated.

\subsubsection{60\%-40\% split}
The training dataset had in average 6,209 annotated entries (after similarity detection) and the testing dataset was composed of 12,338 entries.

Tables~\ref{evaluation6040Basico} and~\ref{evaluation6040Nuevo} summarize the results.  Table~\ref{evaluation6040Basico} shows that our system had the best precision performance if both word and character $n$-grams features are enabled, but All-In-1 was still the best alternative in terms of recall and $F$ for the basic combinations of features. In this case, the precision, recall and $F$ metrics of our system improved versus the baseline by about 27\%, 58\% and 50\%, respectively, after activating all the features. We again attained the best precision, outperforming the best competitor by almost 2\% when all features are activated. Our system ranked second in recall after All-In-1 by a narrow margin of about 3\% again when all features are activated. We remark again that we are not using semantic information from word embeddings.

\begin{table*}[!htbp]
\centering
\caption{\label{evaluation6040Basico}Average evaluation metrics for the basic combinations of features, 60\%-40\% split.}
\small
\begin{tabular}{ccp{8.2cm}ccc}
\hline
\multicolumn{1}{c}{Train} & \multicolumn{1}{c}{Test} & \multicolumn{1}{c}{Features} & \multicolumn{1}{c}{$\mbox{P}_{\mbox{\tiny \bf macro}}$} & \multicolumn{1}{c}{$\mbox{R}_{\mbox{\tiny \bf macro}}$} & \multicolumn{1}{c}{$\mbox{F}_{\mbox{\tiny \bf macro}}$}\\ \hline
\multirow{6}{*}{60\%} & \multirow{6}{*}{40\%}
& Proposed system word $n$-grams & 68.54\% & 29.24\% & 40.97\% \\
& & Proposed system word $n$-grams + char $n$-grams & \bf 94.38\% & 80.42\% & 86.83\%\\\cline{3-6}
& & All-In-1 word $n$-grams & 92.74\% & 89.75\% & \bf 91.12\% \\
& & All-In-1 word $n$-grams + char $n$-grams  \hspace{0.2cm} & 90.98\% & \bf 89.89\% & 90.29\% \\
& & IITP-{\sc cnn}  & 90.54\% & 85.15\% & 87.37\%\\
& & IITP-{\sc cnn}+{\sc rnn} & 89.97\% & 81.42\% & 85.02\%\\
\hline
\end{tabular}
\end{table*}

\begin{table*}[!htbp]
\centering
\caption{\label{evaluation6040Nuevo}Average evaluation metrics of the proposed system for all combinations of features, 60\%-40\% split.}
\small
\begin{tabular}{ccp{8.5cm}ccc}
\hline
\multicolumn{1}{c}{Train} & \multicolumn{1}{c}{Test} & \multicolumn{1}{c}{Features} & \multicolumn{1}{c}{$\mbox{P}_{\mbox{\tiny \bf macro}}$} & \multicolumn{1}{c}{$\mbox{R}_{\mbox{\tiny \bf macro}}$} & \multicolumn{1}{c}{$\mbox{F}_{\mbox{\tiny \bf macro}}$}\\ \hline
\multirow{4}{*}{60\%} & \multirow{4}{*}{40\%}
& Word $n$-grams & 68.54\% & 29.24\% & 40.97\% \\
& & Word $n$-grams + lexica & 85.74\% & 65.42\% & 74.21\%\\
& & Word $n$-grams + lexica + amount + date  & 93.14\% & 68.28\% & 78.79\%\\
& & Word $n$-grams + lexica + amount + date + char $n$-grams & \bf 95.60\% & \bf 87.05\% & \bf 91.12\% \\
\hline
\end{tabular}
\end{table*}

\subsubsection{70\%-30\% split}
The training dataset had in average 6,780 annotated entries (after similarity detection) and the testing dataset had 9,253 entries.

Tables~\ref{evaluation7030Basico} and~\ref{evaluation7030Nuevo} show the results. With the basic features all systems achieved similar results (Table~\ref{evaluation7030Basico}). In this case, the precision, recall and $F$ of our system improved by about 28\%, 57\% and 49\%, respectively after activating all the features. We again attained the best precision and almost matched All-In-1 in terms of recall and $F$ when all features are activated.

\begin{table*}[!htbp]
\centering
\caption{\label{evaluation7030Basico}Average evaluation metrics for the basic combinations of features, 70\%-30\% split.}
\small
\begin{tabular}{ccp{8.2cm}ccc}
\hline
\multicolumn{1}{c}{Train} & \multicolumn{1}{c}{Test} & \multicolumn{1}{c}{Features} & \multicolumn{1}{c}{$\mbox{P}_{\mbox{\tiny \bf macro}}$} & \multicolumn{1}{c}{$\mbox{R}_{\mbox{\tiny \bf macro}}$} & \multicolumn{1}{c}{$\mbox{F}_{\mbox{\tiny \bf macro}}$}\\ \hline
\multirow{6}{*}{70\%} & \multirow{6}{*}{30\%}
& Proposed system word $n$-grams & 67.41\% & 30.67\% & 42.16\% \\
& & Proposed system word $n$-grams + char $n$-grams & \bf 94.37\% & 80.15\% & 86.88\%\\\cline{3-6}
& & All-In-1 word $n$-grams & 92.58\% & \bf 90.01\% & \bf 91.16\% \\
& & All-In-1 word $n$-grams + char $n$-grams \hspace{0.21cm} & 91.35\% & 89.95\% & 90.52\% \\
& & IITP-{\sc cnn}  & 89.33\% & 85.67\% & 87.19\%\\
& & IITP-{\sc cnn}+{\sc rnn} & 90.06\% & 82.14\% & 85.63\%\\
\hline
\end{tabular}
\end{table*}

\begin{table*}[!htbp]
\centering
\caption{\label{evaluation7030Nuevo}Average evaluation metrics of the proposed system for all combinations of features, 70\%-30\% split.}
\small
\begin{tabular}{ccp{8.5cm}ccc}
\hline
\multicolumn{1}{c}{Train} & \multicolumn{1}{c}{Test} & \multicolumn{1}{c}{Features} & \multicolumn{1}{c}{$\mbox{P}_{\mbox{\tiny \bf macro}}$} & \multicolumn{1}{c}{$\mbox{R}_{\mbox{\tiny \bf macro}}$} & \multicolumn{1}{c}{$\mbox{F}_{\mbox{\tiny \bf macro}}$}\\ \hline
\multirow{4}{*}{70\%} & \multirow{4}{*}{30\%} 
& Word $n$-grams & 67.41\% & 30.67\% & 42.16\% \\
& & Word $n$-grams + lexica & 87.91\% & 66.71\% & 75.84\%\\
& & Word $n$-grams + lexica + amount + date & 92.99\% & 70.22\% & 80.01\%\\
& & Word $n$-grams + lexica + amount + date + char $n$-grams  & \bf 95.05\% & \bf 87.51\% & \bf 91.12\% \\
\hline
\end{tabular}
\end{table*}

Table~\ref{evaluationCat7030} shows the performance of our system by {\sc bt} category when all the features are enabled. In general the performance was satisfactory. The worst performance corresponded to the categories with fewer entries in the training set (according to Table~\ref{categoriesDist}). 

\begin{table*}[!htbp]
\centering
\caption{\label{evaluationCat7030}Performance of our system by category with all features enabled, 70\%-30\% split.}
\small
\begin{tabular}{lcccc}
\hline
\multicolumn{1}{c}{Category} & \multicolumn{1}{c}{$\mbox{P}_{\mbox{\tiny \bf macro}}$} & \multicolumn{1}{c}{$\mbox{R}_{\mbox{\tiny \bf macro}}$} & \multicolumn{1}{c}{$\mbox{F}_{\mbox{\tiny \bf macro}}$}\\ \hline
Bank & 92.32\% & 93.08\% & 92.69\% \\
Means of transport & 94.54\% & 90.06\% & 92.24\% \\
Shopping & 89.88\% & 98.07\% & 93.79\% \\
Household expenses & 98.15\% & 91.35\% & 94.62\% \\
Taxes and charges & 96.34\% & 82.18\% & 88.58\% \\
Off-cycle income & 95.32\% & 88.89\% & 91.97\% \\
Payroll & 95.92\% & 88.11\% & 91.76\% \\
Leisure & 95.24\% & 87.36\% & 91.12\% \\
Health, sport and education & 97.01\% & 82.38\% & 89.10\% \\
Insurances & 98.29\% & 94.87\% & 96.54\% \\
Social security, grants and pensions & 98.89\% & 85.00\% & 91.31\% \\
Transfers & 88.84\% & 80.19\% & 84.29\% \\
Business and professional expenses & 89.12\% & 73.22\% & 80.31\% \\
Rentals & 98.00\% & 82.86\% & 89.77\% \\
Others & 97.85\% & 95.02\% & 96.42\%\\
\hline
\end{tabular}
\end{table*}

\subsection{Summary of numerical results}
To evaluate the performance of our system we applied cross-validation in five dataset splits between training and testing subsets (30\%-70\%, 40\%-60\%, 60\%-40\% and 70\%-30\%), to check the robustness of our approach as the sizes of the testing subsets  decreased. In these experiments we added features to the model incrementally to assess their  significance, in the following order: word $n$-grams, lexica, amount, date and character $n$-grams. We compared our system with three competing approaches from the state-of-the-art, All-In-1 and two variants of the IITP method.

The Jaccard similarity detector achieved reductions of training data exceeding 56\% for all splits.

For the 30\%-70\% split, our system attained the best precision. It was inferior to All-In-1 in recall and $F$ unless all features were enabled. If they were,  our system also outperformed its competitors in $F$. For the 40\%-60\% split, our system outperformed the competitors in terms of precision,  recall and $F$ when all features were enabled. It was better in precision even with the basic combination of features. For the 60\%-40\% and 70\%-30\% splits, our system again outperformed the competitors in terms of precision, and the performance gap with All-In-1, in the cases it existed, was reduced. Indeed, our approach is simpler than the competitors, which allowed significant training time reduction.

\section{Use case:{\em CoinScrap}}
\label{secoinscrap}

{\em CoinScrap} launched its mobile app for iOS and Android in November 2016, and since then it has had thousands of downloads. A new version of the application was launched October 2018. It includes  journey improvement for product fulfilment; dynamic ``gamified'' saving rules (e.g. saving when your favourite team wins, or when you take a coffee); and personalised recommendations for financial management.

\begin{figure}[!htbp]
\centering
\includegraphics[width=0.25\textwidth]{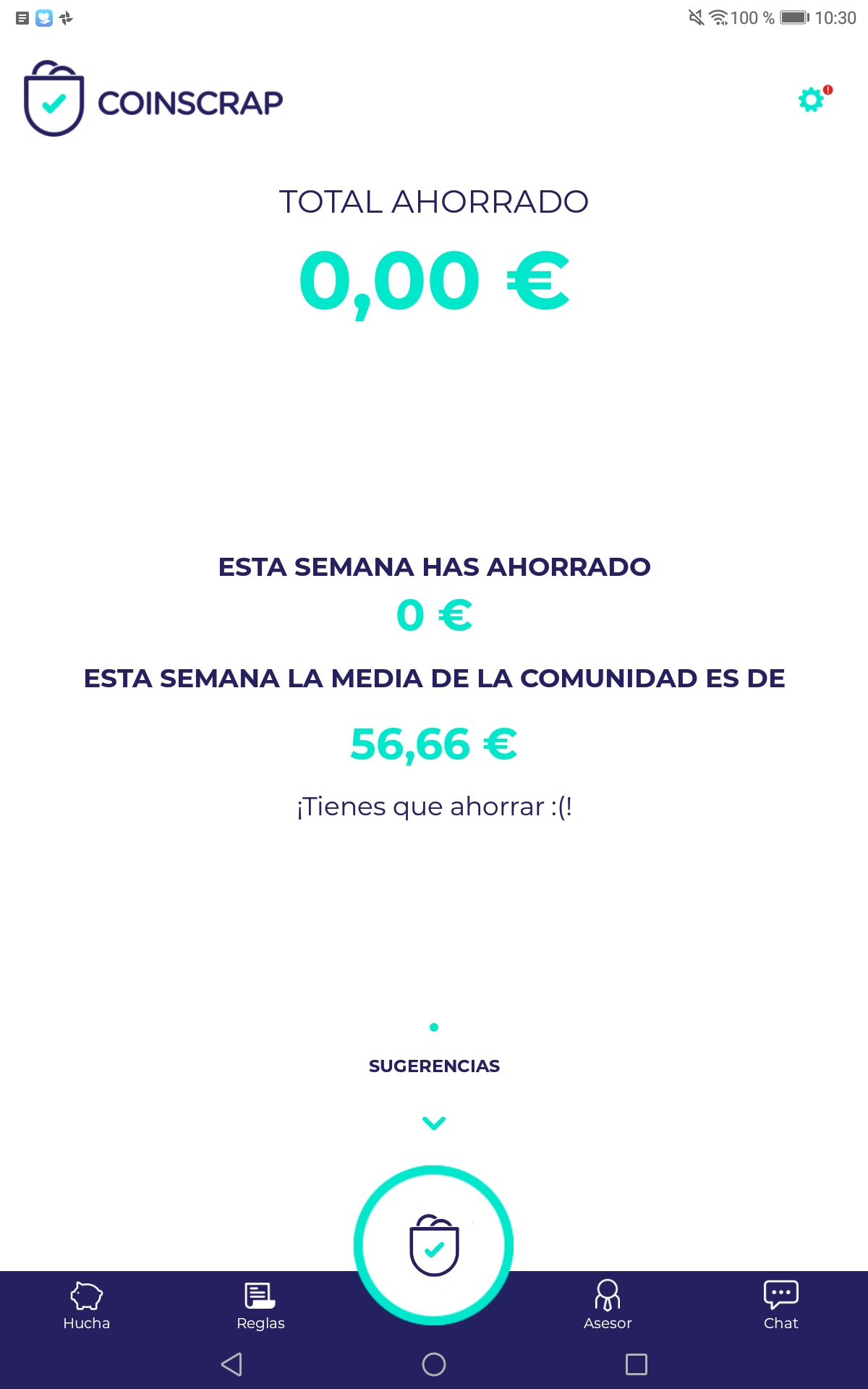}
\caption{\label{coinscrap}{\em Coinscrap} app.}
\end{figure}

The latter rely on  our system to classify {\sc bt} transactions. In this line, {\em CoinScrap} recommends personalized services and products based on  financial necessities and objectives. Figure~\ref{coinscrap} shows an screenshot of the app.

\section{Conclusions}
\label{conclusions}

Compared to normal texts, short texts analysis is challenging due to sparsity, irregularity and real-time data generation. In this paper we describe a short-text {\sc svm} {\sc bt} classification system using a combination of meta-information and linguistic knowledge (by relying on specialized lexica).

Motivated by existing solutions in spam detection, we achieved a significant reduction of training information with a short text similarity detector  based on the Jaccard distance.

Experimental results, by comparing our approach with three state-of-the-art competitors with higher computational complexity, are very promising. Our lexicon feature is crucial to attain high precision, especially if the training dataset is small.

The effectiveness of the proposed system was demonstrated on a real dataset reflecting the activity of real customers of Spanish banks, organized in fifteen different classes including means of transport, shopping, household expenses, taxes, charges and payroll. This labelled dataset is a valuable asset that will be available to other researchers on request.

Our system attained the best precision (which is the most relevant metric in {\sc pfm}) and performed similarly in terms of recall and $F$ if enough features were enabled, especially when the methods were stressed by reducing the training-to-test subset size ratio.

Given the encouraging results in this work, we are currently expanding it to obtain sub-categorisations of the descriptions. Our approach has been put into production in a real {\sc pfm} application, {\em CoinScrap}.

\newpage

\bibliography{mybibfile.bib}{}
\bibliographystyle{IEEEtran}

\EOD

\end{document}